\documentclass[aps,prd,floatfix]{revtex4}
\usepackage{epsf}% Include figure files
\newcommand{\bee}{\begin{equation}}
\newcommand{\ee}{\end{equation}}
\newcommand{\beea}{\begin{eqnarray}}
\newcommand{\eea}{\end{eqnarray}}

\newcommand{\lsim}{\mathrel{\lower4pt\hbox{$\sim$}}
\hskip-12.5pt\raise1.6pt\hbox{$<$}\;}

\def\slash{\!\!\!\!/\,}
\def\Dslash{{D}\slash}

\begin{document}
\title{Recent developments from lattice QCD}

\author{Thomas DeGrand}
\affiliation{
Department of Physics,
University of Colorado,
        Boulder, CO 80309 USA
}

\begin{abstract}
This year lattice QCD has become very public.
A new generation of simulations (including light dynamical quarks) have produced
results which are in close agreement with many ``easy'' experimental quantities, and
precise predictions for quantities which are tests of the Standard Model.
Is QCD over? Reality is somewhat more nuanced. I will try to put the recent results
into context: as in any theoretical calculation,  there are always hidden assumptions!
Hopefully I can give you a feel for some of them.
\end{abstract}

\maketitle

\section{Introduction}

According to the BaBar Physics Book\cite{Harrison:1998yr},
 the purpose of lattice QCD calculations are ``to solve QCD
directly by a numerical simulation.''

Wasn't QCD solved years ago?

Well, not really. For many years people have been able to calculate many non-perturbative quantities
(hadron masses, decay constants) at the 10-20 per cent level, as long as the quark masses
are not too small. That is not really accurate enough to be useful for Standard Model tests.
We think we understand confinement and probably chiral symmetry breaking in the strong coupling
limit of QCD, but there isn't a really convincing story about either of these in the continuum limit
 (like a graduate student
would use in an oral exam). Lattice techniques got us the 10-20 per cent numbers, and will probably
do a lot better in the next few years. There still may not be an answer for the graduate student.

The theory\cite{DeGrand:2003xu}
 behind the BaBar  mission statement goes as follows: One begins with the generating functional
for Green's functions for QCD, regulated with a UV cutoff, with  a set of bare
couplings,  and the bare quark masses
\bee
Z(J) = \int [dA_\mu][d\psi][d\overline \psi] \exp(-S(\bar \psi,\psi,A)). 
\ee
We integrate out the fermions, leaving the gauge action $S_G(A)$ and the fermion determinant,
\bee
Z(J)   = \int [dA_\mu] \det(\Dslash + m)^{N_f}\exp(-S_G(A)) . \nonumber 
\ee
A correlator is measured from (for example)
\bee
C_\Gamma(x,y) = \frac{1}{Z} \int [dA_\mu][d\psi][d\bar \psi] \exp(-S(\bar \psi,\psi,A))
\bar \psi_x \Gamma \psi_x \bar \psi_y \Gamma \psi_y
\ee
which is equal to
\bee
C_\Gamma(x,y) = \int d^4 q \sum_h \frac{|\langle 0| \bar \psi \Gamma \psi | h \rangle|^2}{q^2+m_h^2}
e^{iq(x-y)},
\ee
and masses and matrix elements can be extracted from averages of $C_\Gamma(x,y)$: for example,
\bee
\int d^3 \vec x C(x_0,\vec x;0,0) = \sum_h \exp(-m_h x_0) \frac{|\langle 0 |\bar \psi \Gamma \psi | h \rangle|^2}{2m_h} .
\ee
The lattice comes in when one regulates the action with a UV cutoff which is a mesh of space-time points
(with some lattice spacing $a$), replaces the continuum action with a lattice action (which is a function
of bare parameters defined at the cutoff scale), and replaces the fields by some bare lattice fields.
The bare action and fields are defined so that
any desired symmetries survive discretization. Monte Carlo comes in when one replaces $Z(J)$ by
an ensemble of ``snapshots'' of the gauge fields, where the probability of finding a particular snapshot
is proportional to $\det(\Dslash + m)^{N_f}\exp(-S_G(A)$, and the correlator $C_\Gamma(x,y)$ is
approximated  by
an ensemble average
\bee
\frac{1}{N} \sum_{j=1}^N C_\Gamma(x,y,\{A\}_j) + O(\frac{1}{\sqrt{N}})   .
\ee
The ensemble of snapshots is generated on a computer. All lattice simulations are done at unphysical values
of renormalized constants with a nonzero cutoff, in finite volume. Neglecting quark masses for the moment,
a lattice calculation of a mass ratio will be a ratio of quantities measured
 in units of the cutoff, and  will be equal to
 a cutoff-independent value plus a  sum  of cutoff effects
\bee
\frac{m_1(a) a}{m_2(a) a} = \frac{m_1}{m_2} + O(a) +O(a^2)+\dots .
\ee
 Lattice people talk about ``controlled systematic errors'' when they think that they
can take the volume to infinity, the lattice spacing to zero, and the quark mass to a physical value
(that is, the bare quark mass from cutoff scale $a$ is tuned so that some ratio of hadron masses
takes its experimental value).
The actual route from action to answer soon becomes horribly technical.
Matrix elements of operators with anomalous dimensions have their own complications.

So much for ideology, on to reality. All theoretical calculations have hidden assumptions, and the lattice
is no exception!

Lattice calculations begin by picking a lattice discretization of one's desired theory.
In principle,  constructing lattice actions is no different from the usual particle physics game
that you all play: An author recognizes some interesting IR physics and invents a UV completion for his/her
theory to get it. The interesting IR physics from the lattice was confinement:  A lattice
regulated gauge theory automatically confines if $g(a) \gg 1$. Now the UV cutoff is taken away
(by taking $g(a)$ to zero).
 Maybe the desired IR 
physics will remain. Unfortunately, in addition to the desired IR physics, there will be a set of cutoff
dependent corrections ( $O(p^2 a^2)$ for physics at scale $p$?). These effects are new physics. If they are
not seen in experiment, one must argue that the energy scale for the UV completion is higher than some cutoff.
The only difference between an ordinary particle theorist and a lattice theorist is that the ordinary
theorist might believe that the UV completion corresponds to something real. For us, the lattice
spacing is unphysical, and we know it.
 Nowadays we invent lattice actions which are designed to hide the cutoff effects,
in order to do our simulations at larger values of the cutoff. This is called an ``improvement program.''

Unfortunately, this game has two big problems. First, cutoffs usually don't respect symmetries
unless they are carefully designed. One will typically
have to do some kind of fine-tuning of bare parameters to achieve symmetries in the IR. The
lattice is particularly unfriendly to chiral symmetries, which is not good news for simulations with
small quark mass.

The second problem is that numerical simulation is a big part of a lattice calculation, and computer
resources are always finite. People naturally neglect things that they don't think are important.
It may be hard to correct for this, later. Prime candidates are the extrapolation in
 quark masses,
 the use of the correct number of flavors, or the volume. Dynamical fermions are difficult
because computers can't handle Grassman variables, so they are integrated out at the start. The fermion
determinant is complicated and non-local. For years, lattice calculations have used
the ``quenched approximation,'' in which the number of dynamical flavors is set to zero:
$\det(\Dslash + m)^{N_f} \rightarrow 1$. That's quite an approximation!

\section{The End of Quenching}
\begin{figure}[!thb]
\begin{center}
\epsfxsize=0.6 \hsize
\epsffile{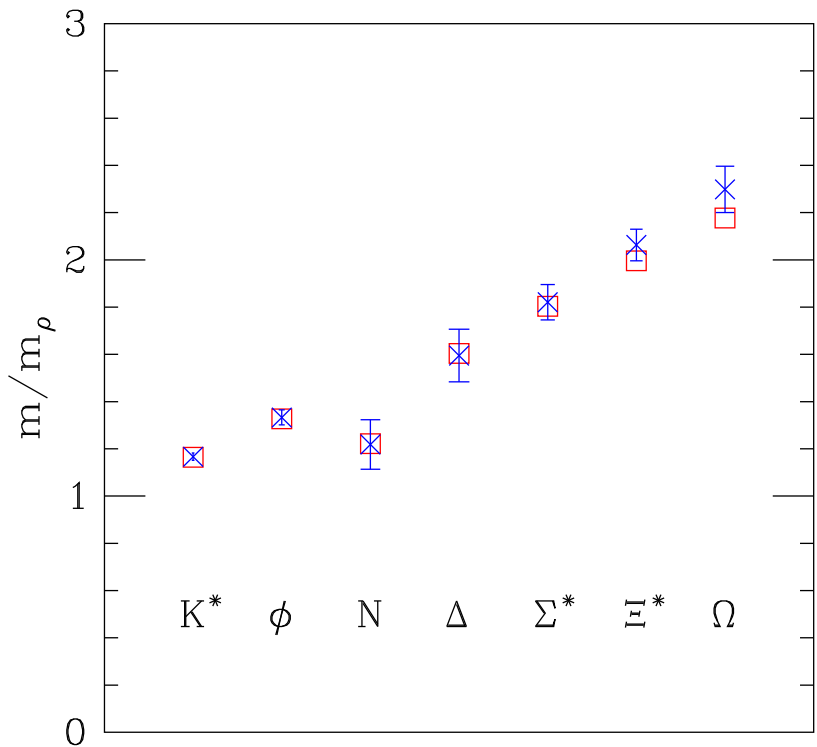}
\end{center}
\caption{
Comparison of quenched results from Ref. \protect\cite{Butler:em}
with experiment.}
\label{fig:wein}
\end{figure}

Lattice phenomenology dates from 1974, with Wilson's discovery of confinement\cite{Wilson:1974sk}
 and 
(a bit later) of chiral symmetry breaking\cite{Greensite:1980hy,Kluberg-Stern:1981wz}
for lattice-regulated gauge theories. The first\cite{Creutz:1979kf}
 Monte Carlo simulations were done in 1979, and the first calculations
of hadron spectroscopy\cite{Hamber:1981zn}
 date from 1981. A standard set of post-dictions (the proton mass) and predictions
($B_K$, $f_B$, $f_D, \dots$) have been part of the lattice menu ever since.
It is a subject characterized by gradual progress, punctuated by little revolutions.
 ``QCD has been solved'' -- several times.
Fig. \ref{fig:wein} shows spectroscopy from 1994 \cite{Butler:em}. You could not ask for anything better.

Of course, standards always go up, and QCD has been un-solved several times, too.

To view the past, we again visit the BaBar Physics Book (fall 1998), which has an appendix about lattice predictions.
At that time, $f_D$ and $f_B$ had about a 20 MeV statistical uncertainty, $B_B$ was a 5-10 \% number, and
$B_K=0.61(6)$.
The  book did not quote a big systematic uncertainty: these calculations were
done in quenched approximation. That apparently matters a lot for some quantities. Back then
$f_{D_s}\simeq 220(30)$ MeV. A recent calculation\cite{Simone:2004fr}
 with 2+1 flavors gives $f_{D_s} = 263(+5,-9)(24)$ MeV
(the two error bars are statistics and estimates of matching to the continuum),
 which is certainly not the same number. This illustrates the size of a ``quenching systematic.''

The quenched approximation has many of the ingredients of successful hadron phenomenology.
Quarks are confined (with a linear confining potential if they are heavy).
Chiral symmetry is spontaneously broken.
In it, all states are (at first glance) infinitely narrow,
 because $q \bar q$ pairs cannot pop out of the vacuum.
One might also try to ``justify'' the quenched approximation by an appeal to the quark
model: in the quenched approximation, all mesons are $q \bar q$ pairs, and all baryons
are $qqq$ states.
This also appears to be rather similar to the large-$N_c$ limit of QCD.

The best way to see what is going on in the quenched approximation is to
consider the low energy limit of QCD. Do not think about quarks and gluons, but
in terms of an effective field 
theory of QCD, described by a chiral Lagrangian in which the would-be
Goldstone bosons are fundamental fields. These Lagrangians have
a set of bare parameters (quark masses, $f_\pi$, the quark condensate $\Sigma$, $\dots$).
As far as the chiral Lagrangian is concerned, these are fundamental parameters.
As far as QCD is concerned, one could compute these parameters from first
principles (for example, $f_\pi m_\pi = \langle 0 |\bar \psi \gamma_0\gamma_5
\psi|\pi\rangle$),
this would fix the parameters of the chiral Lagrangian, and then one could throw away the
 lattice and compute low energy physics using the
chiral Lagrangian. Quenched QCD and QCD with nonzero flavor numbers are different theories
and their low energy parameters will be different. But there is more.
In full QCD the eta prime is heavy and can be
decoupled from the interactions of the ordinary Goldstone bosons. In quenched QCD
the eta prime is not really a particle. The would-be eta prime
gives rise to ``hairpin insertions'' which pollute essentially all predictions.

\begin{figure}[!thb]
\begin{center}
\epsfxsize=0.7 \hsize
\epsffile{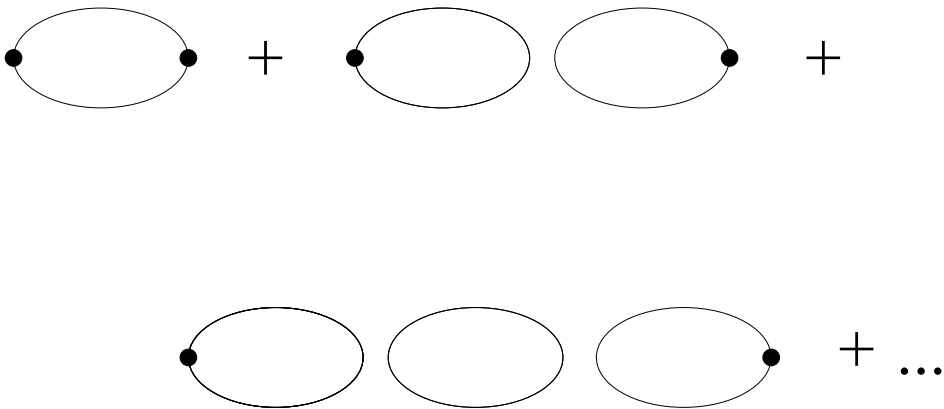}
\end{center}
\caption{
The eta-prime propagator in terms of a set of annihilation graphs
summing into a geometric series
to shift the eta-prime mass away from the mass of the flavor non-singlet
pseudoscalar mesons. In the quenched approximation, only the first two
terms in the series survive as the ``direct'' and ``hairpin'' graphs.
}
\label{fig:hairpin}
\end{figure}

Let's consider the eta prime channel in
full QCD and quenched QCD.
In ordinary QCD, the eta prime
propagator includes a series of terms in which the flavor singlet
$q\bar q$ pair annihilates into some quarkless state, then reappears, over and over.
This is shown in Fig. \ref{fig:hairpin}.  The eta prime propagator is
\bee
\eta'(q) = C(q) -H_0(q) + H_1(q) + \dots
\ee
where $C(q) = 1/d$, $d=q^2+m_\pi^2$, is the ``connected'' meson propagator, the
same as for any other Goldstone boson. $H_n$ is
the $n$th order hairpin (with $n$ internal fermion loops). Assuming that each vertex is a constant $V$
allows us to sum the geometric series
\bee
\eta'(q) = \frac{1}{ q^2+m_\pi^2 +V}
\label{eq:etaprimepole}
\ee
and generate a massive eta prime.

However, the quenched limit is different -- there are no loops. In that case
(of course)
\bee
\eta'(q) = {1\over d}  - {1\over d} V {1\over d}.
\label{eq:etaq}
\ee
In the eta prime channel there is an ordinary (but flavor singlet) Goldstone
boson and a   new contribution--a double-pole ghost (negative norm)
state. In the $N_c=\infty$ limit, the double pole decouples, but finite $N_c$ quenched QCD
remains different from finite $N_c$ full QCD. The limits of large $N_c$ and
quenching don't commute.

Where the eta-prime comes in  is in the
calculation of corrections to tree-level relations\cite{BG,Sharpe:1992ft}.
 These are typically dominated
by processes with internal Goldstone boson loops,
contributing terms like
\bee
\int d^4 k G(k,m) \simeq ({m \over {4\pi}})^2 \log({{m^2}\over{\Lambda^2}})
\ee
(plus cutoff effects). The eta-prime hairpin can appear in these loops,
replacing $G(k,m) $ by $ -G(k,m) V G(k,m)$ and altering the chiral logarithm.
Thus, in a typical observable, with a small mass expansion
\bee
Q(m_{PS}) = A(1 + B{ m_{PS}^2 \over f_{PS}^2 }\log m_{PS}^2 )+ \dots
\label{eq:eq001}
\ee
quenched and $N_f=3$ QCD
can have different coefficients (different $B$'s in Eq. \ref{eq:eq001}), seemingly
randomly different.  (Quenched $f_\pi$ has no chiral logarithm while it does in full QCD,
the coefficient of $O_+$, the
 operator measured for $B_K$, is identical in quenched and full QCD,
etc.)
Even worse, one can find a different functional form. For example, the
relation between pseudoscalar mass and quark mass in full QCD is
\bee
m_{PS}^2 = Am_q(1 + {{m_{PS}^2}\over{8\pi^2f^2}}\log(m^2/\Lambda^2)]+\dots.
\ee
In quenched QCD, the analogous relation is
\bee
(m_{PS})^2/(m_q) = A [1 - \delta (\ln ( m^2/\Lambda^2) +1 )] + \dots
\label{eq:log}
\ee
where $\delta=V/(8\pi^2  N_c f_\pi^2)$ is expected to be about 0.2
using the physical $\eta'$ mass.
This means that $m_{PS}^2/m_q$ actually diverges in the chiral limit!

While there are some lattice observations of Eq. \ref{eq:log} behavior (with noisy fits to $\delta$),
what has really happened is a crisis in confidence for phenomenology. People want to 
extrapolate their simulations (run at unphysically heavy quark masses) to the chiral limit.
The best way to do that is to use an effective chiral Lagrangian to predict the
quark mass dependence. But if the chiral Lagrangian which describes quenched QCD is different
from the one which describes $N_f=3$ QCD, what are you supposed to do? (And for that matter,
if quenched QCD and $N_f=3$ QCD are really different theories, how can you be sure that matching
one or a few parameters means that other quantities will match?)

And since people think that they can do simulations with dynamical quarks, why bother with the
quenched approximation?

We are in the middle of an exciting and peculiar time  in lattice QCD, with two essentially
uncoupled developments.
One of them is much more mature from the point of simulations, and it is the one which has
gotten most of the publicity lately: these are simulations with three flavors of light
dynamical fermions. The other development is the discovery of lattice fermion actions which
have exact chiral symmetry at nonzero lattice spacing.

\section{Lattice Chiral Symmetry in a Nutshell}
A naive discretization of the Dirac operator $\bar \psi \gamma_\mu p_\mu \psi \rightarrow
\bar \psi \gamma_\mu \frac{1}{a}\sin(p_\mu a)\psi$ ``doubles'' the spectrum; all the modes
with at least one $p_\mu \simeq \pi/a$ are as light as the one near $p=0$, and the chiral charges
of all the doublers plus the $p=0$ mode cancel. The Nielsen-Ninomiya\cite{Nielsen:1980rz} theorem encodes
a dilemma: no lattice fermion action can be quadratic, well-behaved, have a conserved local
axial charge $Q$ which is quantized, without having an equal number of left-handed and right-handed 
fermions for each eigenvalue of $Q$. Thus the ``classic'' division of lattice fermions into Wilson-like,
which have no doublers but explicitly break chiral symmetry with higher dimensional operators,
or staggered fermions, which double but preserve a relic of chiral symmetry. An exact
transformation and decimation  to one-component fermions
living on the sites of a hypercube  shrinks the multiplicity-16 naive fermion into a
 multiplicity-4 staggered fermion,
and leaves a $U(1)\otimes U(1)$ chiral symmetry.

However, if something is really forbidden, they don't pass laws against it, and smart people
invented a number of ways to evade the Nielsen-Ninomiya theorem. Back in 1982, Ginsparg
and Wilson\cite{Ginsparg:1981bj} used renormalization group ideas to propose a modification of
chiral symmetry; they could not provide an explicit
example of an action and the idea was forgotten. In the early '90's Kaplan\cite{Kaplan:1992bt} and
 later Shamir\cite{Shamir:1993zy}
studied QCD in five-dimensional worlds. The sound bite is very familiar: we (and the chiral
fermion) live on a brane or a boundary in a higher dimensional world. A transfer matrix version of this
idea, to give a chiral four-dimensional fermion,
 was developed at the same time by Narayanan and Neuberger\cite{Narayanan:1994gw}.
(Quenched) simulations with domain wall fermions began in 1996. The bulk of the community not doing these things
woke up in 1997 when the Ginsparg-Wilson was rediscovered, and Luscher\cite{Luscher:1998pq} gave us a modified rule
for a chiral transformation involving the Dirac operator $D$ itself
\bee
\delta \psi = \epsilon\gamma_5(1 + aD) \psi; \quad\quad \delta \bar \psi = \epsilon\bar\psi \gamma_5
\ee
which encodes the
Ginsparg-Wilson relation,
\bee
\{\gamma_5, D \} = a  D \gamma_5  D.
\ee
This is a magic formula, which guarantees that all chiral Ward identities are satisfied on the
lattice at nonzero lattice spacing, up to contact terms. The index theorem is exact for the overlap:
topology can be computed by counting fermionic chiral zero modes.
The overlap of  Neuberger and Narayanan is an exact\cite{Neuberger:1997fp}
 realization of the Ginsparg-Wilson relation, using any nonchiral Dirac operator $d$, it is
\bee
D_{ov} = R_0(1 + \frac{ d - R_0}{\sqrt{|d - R_0|^2}} )
\ee
With it, quenched simulations with the overlap began a year or two later.

\section{Staggered Fermion Phenomenology}
\begin{figure}[!thb]
\begin{center}
\epsfxsize=0.5 \hsize
\epsffile{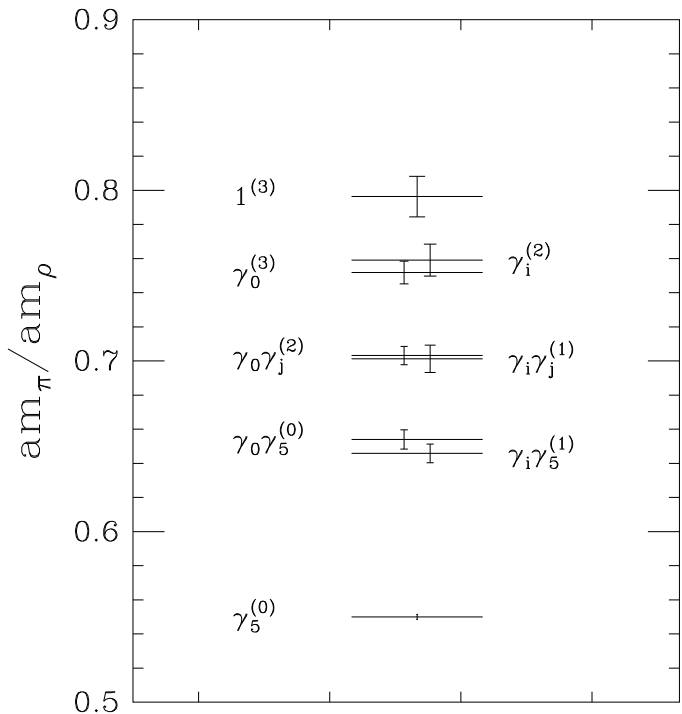}
\end{center}
\caption{An example of flavor or taste symmetry breaking in an improved
staggered action. The different $\gamma$'s are a code for the various
pseudoscalar states.
Data are from Ref. \protect\cite{Orginos:1999cr}. For an explanation of
the splitting, see Ref. \protect\cite{SHARPLEE}.}
\label{fig:flav}
% FIGURE_FILE clover.ax
\end{figure}

With essentially zero overlap with the last section, since 1987  members of the MILC collaboration
have been doing simulations of QCD with two and three flavors of ever lighter dynamical
staggered  fermions.
By 2001 they were up to lattice sizes of $L=2 \ fm$, and down to
 a strange quark at its physical value, and 
 non-strange quarks of about 20 MeV. A major problem with staggered quarks is that interactions
mix and split hadrons made of the four ``tastes'' of a single staggered flavor, producing
a pattern reminiscent of the effect of crystal fields on atomic spectral lines
(See Fig. \ref{fig:flav} for an example.) The MILCmen\cite{Orginos:1999cr}
 redesigned the gauge connection of the staggered
fermion to reduce this to a small value, so that it made sense to talk about pions and kaons
as separated collections of states.

Their next crucial breakthrough was in the data analysis:
The key to doing this was provided by the Sharpe and Lee\cite{SHARPLEE}
analysis of taste mixing,
whose construction of a low energy chiral effective theory including explicit
taste-breaking interactions  predicted the
degeneracies shown in Fig. \ref{fig:flav}. Their work was generalized by Aubin
and Bernard\cite{Aubin:2003uc}. One writes down a low energy effective field theory for the
Goldstones ($\Sigma$ is the usual exponential of the particle fields,
 traces over the 16 taste-product $q\bar q$  bilinears of
each staggered flavor) and the
 Lagrangian is
\begin{eqnarray}\label{eq:final_L}
        {\mathcal{L}} & = & \frac{f^2}{8} {\rm{Tr}}(\partial_{\mu}\Sigma
        \partial_{\mu}\Sigma^{\dagger}) -
        \frac{1}{4}\mu f^2 {\rm{Tr}}( {\mathcal{M}} \Sigma+{\mathcal{M}}\Sigma^\dagger) \\
        &  & + \frac{2m_0^2}{3}(U_I + D_I + S_I + \cdots)^2 + a^2 {\mathcal{V}},
\end{eqnarray}
where the $m_0^2$ term weighs the  analog of the flavor singlet $\eta'$.
(The ``$I$'' subscripts display that this involves the
taste singlet term for each flavor.)
 The $a^2 {\mathcal{V}}$
term is the taste-breaking interaction, a sum of terms quadratic in  $\Sigma$ with
various taste projectors, parameterized by six coefficients (only one is big).
Now one computes ``any'' desired quantity with this Lagrangian, typically to
one loop, as a function of quark masses and all other coefficients.
Parameters of Nature are determined when mass-dependent Monte Carlo data is fit
to this functional form.  For example, a one-loop fit to $m_{PS}/(m_1+m_2)$ for the
true would-be Goldstone boson made of  quarks of mass $m_1$ and $m_2$
 would involve $\mu$, $f$, two Gasser-Leutwyler parameters, three otherwise
unconstrained lattice parameters, and involves chiral logarithms whose arguments
are all the observed pseudoscalar masses. Fits to $f_\pi$ or $f_K$ are similar.

\begin{figure}[!thb]
\begin{center}
\epsfxsize=0.7 \hsize
\epsffile{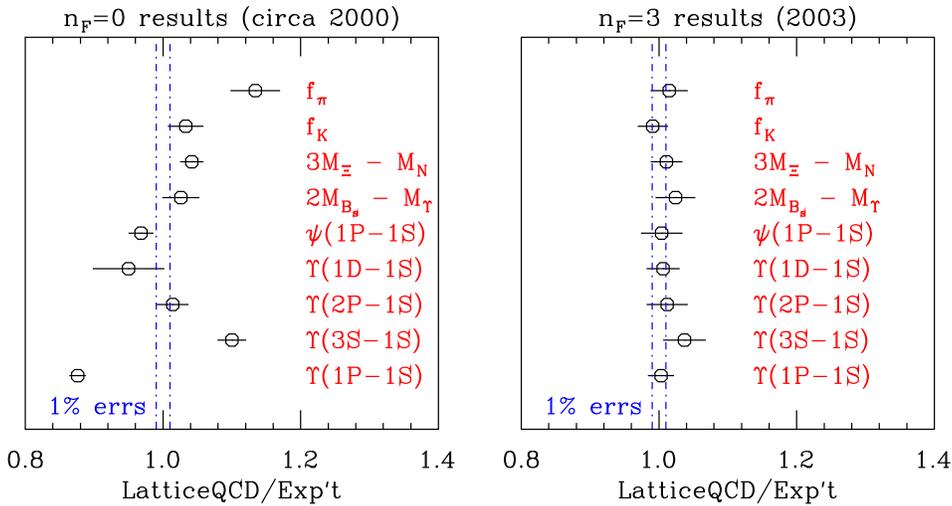}
\end{center}
\caption{
Comparison of quenched results with results from simulations with 2+1 flavors
of staggered fermions, from Ref. \protect\cite{Davies:2003ik}}.
\label{fig:golden}
\end{figure}

Last year they were joined by other collaborations who used their configurations for
backgrounds for  other QCD phenomenology. In Ref. \cite{Davies:2003ik}  they presented 
results for a variety of post-dictions, which certainly set a new standard for claimed
precision: Fig. \ref{fig:golden} shows their results.

\begin{figure}[!thb]
\begin{center}
\epsfxsize=0.5 \hsize
\epsffile{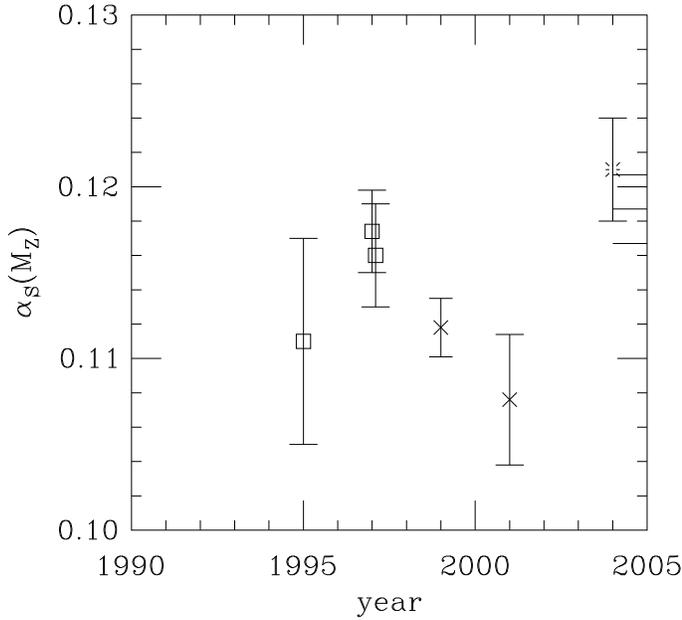}
\end{center}
\caption{Lattice calculations of $\alpha_{\overline{MS}}(M_Z)$ vs. year of publication. The burst
is the recent 2+1 flavor result of Ref. \protect\cite{Davies:2003ik}. Squares are earlier
staggered simulations, mostly with two flavors of dynamical simulations,
and the crosses are $N_f=2$ simulations with clover or Wilson fermions. The horizontal lines
along the right edge show the one-sigma PDG average.}
\label{fig:alphas}
% FIGURE_FILE clover.ax
\end{figure}

Since then, they and their collaborators have gone on to do
\begin{itemize}
\item{
The strong coupling constant at the Z-mass: see Fig. \ref{fig:alphas}
}
\item{
Form factors for semileptonic B and D meson decay (See Ref. \cite{Gray:2004hd}).
}
\item{
Marciano\cite{Marciano:2004uf} has proposed using a lattice calculation of $f_K/f_\pi$ to fix $V_{us}$. The
present MILC data give $V_{us}=0.2236(30)$; the error is compatible with other determinations
of $V_{us}$ and can be shrunk by better lattice simulations
}
\item
{Blum \cite{Blum:2004cq} is using MILC data to compute the hadronic contribution to muon $(g-2)$ from first principles
}
\item{
MILC\cite{Aubin:2004fs} has determined the strange and nonstrange quark masses from a fit to spectroscopy: 
$m_s(\overline{MS},\mu=2$ GeV$)=76$ MeV, the nonstrange average mass 2.8(4) MeV. The up quark is not
massless, by many standard deviations.
}
\end{itemize}
and much more.

There is one problem with the MILC results, however. Recall that a single staggered flavor corresponds
to four continuum ``tastes.'' To get the correct flavor weighting of the determinant, MILC
takes the quarter root of the staggered determinant so that $\det(D_{stagg})^{1/4}$
approximates
$\det(D_{1-flavor})$. If we had a theory of 4 degenerate fermions, each with its Dirac operator
$D_1$, then one could define an operator $D_4 = (D_1)^4$ and
then $\det(D_4)^{1/4}=\det D_1$. But gluons introduce taste-breaking terms among the four tastes,
and it is not clear if there is an analog of $D_1$ for staggered fermions, which
 is theoretically well behaved. If it exists, it would be undoubled and chiral.
It would, therefore,
collide with the Nielsen-Ninomaya theorem unless its chiral properties are unusual. 

Lattice simulations treat valence quarks and sea quarks differently. Basically, valence
quarks (the ones attached to the external sources) do not have any quantum numbers
(other than their mass and spin). One computes
classes of Feynman diagrams on the lattice, then re-weights them with global symmetry indices
 and bundles them
together. (For example, the same propagators are used for the quark and the antiquark in
a mass-degenerate meson). In staggered fermions, one uses a single flavor of
staggered fermions, with its four tastes, and computes correlation functions in which
the sources project (nearly) onto the same initial and final taste. The quark
could hop temporarily into a different taste state as it propagates across the lattice
(this would happen by emitting and absorbing hard gluons), but this is just
cutoff scale physics which contributes $O(a^2 g^2)$ scale violations. It is the sea quarks which
are the problem, and it is a peculiar exchange of limits problem: if taste symmetry were exactly
restored below some lattice spacing, then the spectrum would be exactly degenerate,
and the fractional determinant would correctly count the eigenvalues of a single flavor.
But that is not what happens, taste symmetry is only restored in the continuum limit.

The whole business is very puzzling and unresolved. Some people who worry about this believe that
there is no local action whose determinant is equal to $\det(D_{stagg})^{1/4}$, so that the theory
which is being simulated is not a legitimate quantum field theory. Being nonlocal is bad: there
is no possibility of a renormalization group, because short distance physics cannot be integrated out.
This means universality is lost.

Bunk,  Della Morte, Jansen and Knechtli \cite{Bunk:2004br}
did a direct study of whether $D_{stagg}^{1/4}$ (actually a form of
 $(D_{stagg}^\dagger D_{stagg})^{1/2}$)
was local. It was not. However, saying that two matrices have identical determinants is not the same
as saying that the matrices are identical. All one needs is to find one local $D$, the fact others are
nonlocal is not important.

A number of people\cite{Follana:2004sz,Follana:2004mg,Wong:2004nk}
 have done comparisons of staggered fermions with overlap fermions,
with various choices for discretizations of the gauge-fermion connections. They compare the
spectrum of Dirac eigenmodes in various circumstances.
Staggered fermions do not have  zero modes in the presence of instantons, but
 people find that they can tune these actions to make the modes more and more small and degenerate.
To me, this is more a statement about the properties of the valence quarks than an exact result about
determinants.

D\"urr and Hoelbling\cite{Durr:2004ta}
 have done simulations of the Schwinger model with fractional powers
of staggered quarks and with overlap quarks. They ascribe the differences they see to cutoff dependence
of the staggered quarks, but do not see a smoking gun of anything going obviously wrong.

Finally, there has been a spate of activity with free field theory.
Maresca and Peardon \cite{Maresca:2004me}  have constructed local actions for free fermions
 whose dispersion relation is equal to that for staggered fermions and whose determinant is equal to
$\det(D_{stagg})^{1/4}$. They must impose a Ginsparg-Wilson type of chiral rotation on their
construction to get locality for this transformation.
Adams\cite{Adams:2004mf} has proposed a Wilson-type action with the same determinant.
Shamir\cite{Shamir:2004zc}
 has shown that under RG transformations the free staggered action blocks into
an action with a quadrupled spectrum (in the limit of infinitely many blocking steps). 
If one could construct such an action\cite{Bietenholz:1996qc} and use it in simulations, all would be well,
but that is not what is done in practice: the fractional root is taken first. 
 It is unknown if any of these constructions can be extended
to the interacting theory.

MILC's fits to the staggered chiral Lagrangian include the effective number of sea quarks
as a free parameter. They have done fits freeing it, and find\cite{Aubin:2004fs}
 (for various observables) 1.2 to 1.4
``fit flavors'' (with an uncertainty of about 0.2)  where 1 is desired.

None of these studies  constitutes a demonstration for or against the $\det^{1/4}$ trick. 
For the present, MILC data must dominate any
phenomenological analysis which needs a hadronic matrix element to constrain a Standard Model
parameter. It is clearly better than a quenched calculation, and their quark masses are the smallest
ones being simulated.
No other
formulation of fermions allows simulations at such small quark masses nor large volumes:
it is hard to get below $m_{PS}/m_V =0.6$. 

\section{Conclusions}
Lattice people are hard at work. The MILC program is a major part of American lattice QCD. Many groups are
using their configurations to do phenomenology including the effects of dynamical fermions.
MILC is doing upgrades -- dropping the quark mass and the lattice spacing, raising the volume.
These are the best lattice numbers for phenomenologists to use to date. But the problem remains:
is the $\det^{1/4}$ trick a controlled approximation, or not?

Simulations with dynamical chiral fermions are just beginning.
 The RBC collaboration\cite{Aoki:2004ht}
 is doing
production runs with big lattices with $N_f=2$ flavors
of domain wall fermions. Dynamical overlap simulations are theoretically clean and
beautiful, but remote. Only a few visionary (?) people are playing with them\cite{Fodor:2003bh}.
For you at this conference, the most interesting message might be that such methods
 are beginning to appear;
one might be able to cleanly address interesting questions about chiral fermions with them.

Presumably QCD will be solved a few more times before we retire.

%%%%%%%%%%%%%%%%%%%%%%%%%%%%%%%%%%%%%%%%%%%%%%%%%%%%%%%%%%%%%%%%%%%%%%
\section*{Acknowledgments}
%%%%%%%%%%%%%%%%%%%%%%%%%%%%%%%%%%%%%%%%%%%%%%%%%%%%%%%%%%%%%%%%%%%%%%

This work was supported by the US Department of Energy.

\end{document}